\documentclass{osa-article}

\journal{osajournal}

\setprjcopyright

\articletype{Research Article}

\begin{document}

\title{Efficient inverse design and spectrum prediction for nanophotonic devices based on deep recurrent neural networks}

\author{Ruoqin Yan, Tao Wang\authormark{*}, Xiaoyun Jiang, Qingfang Zhong, Xing Huang, Lu Wang, Xinzhao Yue, Huimin Wang, and Yuandong Wang}

\address{Wuhan National Laboratory for Optoelectronics, Huazhong University of Science and Technology, Wuhan 430074, People's Republic of China}

\email{\authormark{*}wangtao@hust.edu.cn} 



\begin{abstract}
In recent years, the development of nanophotonic devices has presented a revolutionary means to manipulate light at nanoscale. Recently, artificial neural networks (ANNs) have displayed powerful ability in the inverse design of nanophotonic devices. However, there is limited research on the inverse design for modeling and learning the sequence characteristics of a spectrum. In this work, we propose a novel deep learning method based on an improved recurrent neural networks to extract the sequence characteristics of a spectrum and achieve inverse design and spectrum prediction. A key feature of the network is that the memory or feedback loops it comprises allow it to effectively recognize time series data. In the context of nanorods hyperbolic metamaterials, we demonstrated the high consistency between the target spectrum and the predicted spectrum, and the network learned the deep physical relationship concerning the structural parameter changes reflected on the spectrum. Moreover, the proposed model is capable of predicting an unknown spectrum based on a known spectrum with only 0.32\% mean relative error. We propose this method as an effective and accurate alternative to the application of ANNs in nanophotonics, paving way for fast and accurate design of desired devices.
\end{abstract}

\section{Introduction}
Metamaterials and metasurfaces provide several means to manipulate light at nanoscale, which can help realize numerous functionalities, such as cloaking\cite{cloaking1, cloaking2}, enhancing spontaneous emission\cite{enhancing1}, high sensitivity biosensing\cite{sensing1,sensing2}, and negative refractive index\cite{NFI1,NFI2}. Owing to their unique electromagnetic properties, these devices have attracted significant attention from researchers. Nowadays, nanophotonic devices, such as metamaterials and metasurfaces, increasingly rely on complex structures to realize sophisticated functionalities. Designing these devices efficiently is a challenge faced by researchers. Conventional design approaches are usually implemented by trial-and-error methods. A typical design process starts with a set of random parameters and calculates its response by time-consuming electromagnetic methods. Then, the results are compared with the target response, and a modified parameter is computed to update the design. This process depends heavily on the designer's experience and often requires extensive simulations before obtaining a desired design. By contrast, a more direct approach, inverse design, has demonstrated enormous superiority to other approaches. In general, commonly used inverse methods range from genetic algorithms\cite{GA} and adjoint-based topology optimization\cite{top} to particle swarm optimization\cite{PSO}. These methods have been used to design nonlinear photonic switches\cite{adjoint_nonlinear}, dual-mode demultiplexer\cite{adjoint_wang2020inverse}, and other metadevices with exceptional performance. However, since each iteration is computationally expensive, it results an intractable scan of multidimensional parameter spaces within a tolerable time.

To improve the design efficiency without extensive computation of Maxwell’s equations, recently, artificial intelligence methods, especially deep neural networks (DNNs), have been applied in the design of nanophotonic devices\cite{AI1}. In general, inverse design by several deep learning (DL) algorithms including fully connected neural networks (FCNNs)\cite{FCNN1,FCNN2}, convolutional neural networks (CNNs)\cite{CNN1,CNN2} and generative networks (GNs)\cite{GAN1} have achieved excellent results. For designing a regular device, FCNNs and CNNs are the more commonly used methods. In many case of inverse designs, the basic DL algorithm, FCNNs, can learn the relationship between the physical responses and structure. Compared with FCNNs, CNNs are better in extracting spatial features of spectrum and performing analysis on them, resulting in significantly better performance in inverse design\cite{CNN2}. In the typical inverse design, the main design target is a spectrum, a typical sequential data. However, the sequence characteristics of the spectrum, an important feature, have been ignored in the aforementioned DL model. To the best of our knowledge, there is limited research on the inverse design for modeling and learning the sequence characteristics of spectrum. Another major DL network, recurrent neural networks (RNNs) are quite powerful in modeling sequence data such as time series and are among the most widely used models for natural language processing (NLP) today. RNNs feature nontrivial ability in learning sequential data such as speech recognition\cite{RNN_speech},  image generation\cite{RNN_image}, language processing\cite{RNN_language_processing}, and sentiment classification\cite{RNN_sentiment_classification}, and have achieved impressive results.

In this paper, we propose a novel DL method using improved RNNs to extract the sequence characteristics of a spectrum and achieve inverse design and spectrum prediction. The proposed RNNs method can accurately design geometric structure of a desired spectrum response with ease and can capture the deep physical relationship between structure and its response. Furthermore, similar to stock forecast, the model is capable of predicting unknown spectrum based on known spectrum with only 0.32\% mean relative error. Moreover, compared with the parameters of previously proposed inverse design models, such as FCNNs and CNNs, the parameters of RNNs based model are two orders of magnitude fewer. The inverse design model based on the RNNs method is computationally efficient and easy to optimize, implying unparalleled advantages for designing nanophotonic devices.

\begin{figure}[h!]
\centering\includegraphics[width=13cm]{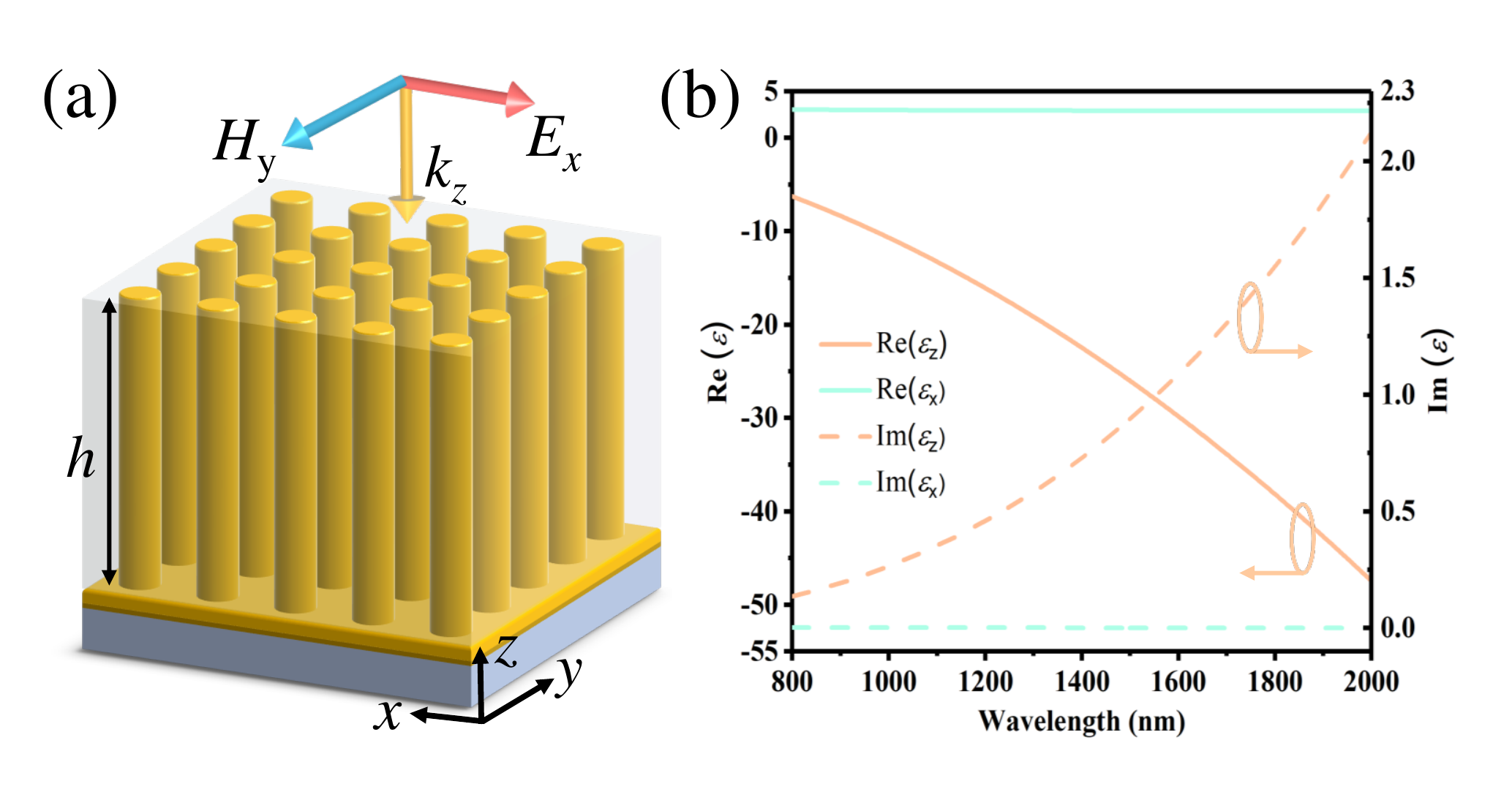}
\caption{(a) Schematic of the NHMM where nanorods are embedded in a porous alumina ($\rm Al_{2}O_{3}$) host matrix resides on a multilayered substrate. The radius (height) of the nanorod is $r$ ($h$) and the arrangement period is $p$, generating a nanorod volume filling fraction $f=\pi {{r}^{2}}/{{p}^{2}}$. (b) Calculated effective permittivity for the metamaterial with the $f$ = 0.234. The real (imaginary) part is shown by the solid (dashed) line.}
\label{fig:1model}
\end{figure}
\section{Modeling and discussion}
To demonstrate the applicability of the RNNs method for nanophotonic design, we use a nanorod hyperbolic metamaterial (NHMM) as an example. HMMs, which feature hyperbolic dispersion because one of their principal components of electric or magnetic effective tensor has the opposite sign to the other two, are unusual electromagnetic metamaterials, since their peculiar properties include enhancing spontaneous emission\cite{enhancing1}, ultrabroadband and anisotropic absorption\cite{HMM-Ultrabroadband,HMM-anisotropic}, negative refraction\cite{HMM-NFI}, biosensing\cite{sensing1,sensing2} and subdiffraction-limited nanolithography\cite{HMM-lithography}. The diagram of the NHMM structure is shown in Fig. \ref{fig:1model}. NHMM can be prepared by embedding array of gold (Au) nanorods in a porous alumina ($\rm Al_{2}O_{3}$) host matrix of thickness $h$. The NHMM sits on a multilayered substrate, which comprises a 1 mm thick glass slide, a 5 nm thick Cr base adhesion layer, and a 30 nm thick Au film acting as the working electrode for the electrochemical growth. NHMMs usually contain three structural parameters, namely the height and radius of the nanorod and lattice period. These parameters are critical to the performance of the device. In this work, two different functional models based on RNNs are built to design NHMMs. First, we implement the inverse design model, that is, given the target spectrum, the corresponding structural parameters can be obtained accurately and quickly. Second, the spectral prediction model can accurately predict unknown spectral data based on known spectral data.
\begin{figure}[h!]
\centering\includegraphics[width=13cm]{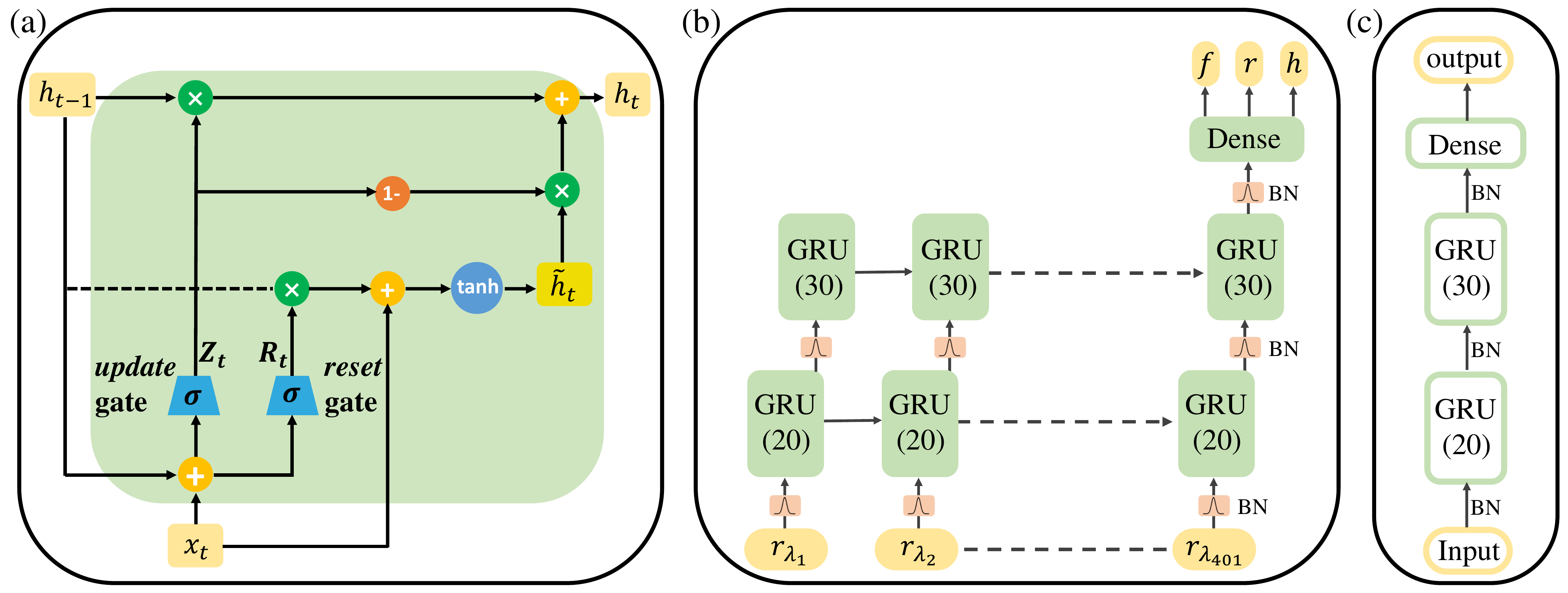}
\caption{(a) Proposed hidden activation function. The update gate $Z_{t}$ selects whether the hidden state is to be updated with a new hidden state  $h_{t}$. The reset gate $R_{t} $ decides whether the previous hidden state is ignored. See Equations. (\ref{R})–(\ref{ht}) for the detailed equations of $R_{t}$, $Z_{t}$, ${\tilde{h}}_{t}$ and $h_{t}$. (b) Architecture of the GRUs inverse design model. This network contains an input layer, two hidden layers, and an output layer. Batch normalization (BN) is used between layers. The reflectance corresponding to 401 wavelength points are input to the network in order. The first and second hidden layers contain 20 and 30 memories in each unit. The last moment activation value of the second hidden layer passes through a dense layer and outputs the structural parameters. (c) Simplification of (b).}
\label{fig:2GRU model}
\end{figure}
\subsection{Gated recurrent units (GRUs) model}
In 2014, Kyunghyun Cho et al proposed GRUs\cite{RNN_GRU}, which is a gating mechanism in RNNs, motivated by the long short-term memory (LSTM) unit\cite{RNN_LSTM} but much simpler to implement and robust to compute. GRUs have an advantage in solving the vanishing gradient, a common issue with RNNs. The graphical depiction of the GRUs is shown in Fig. \ref{fig:2GRU model}(a). First, let us discuss how the recurrent unit of the $t$-th timestep is calculated. Each unit contains two activation gates, the reset gate $R_{t}$ is computed using
\begin{equation}\label{R}
  {{R}_{t}}={\rm \sigma} \left( {{W}_{xr}}{{x}_{t}}+{{W}_{hr}}{{h}_{t-1}}+{{b}_{r}}\right),
\end{equation}
where $\rm \sigma$ represents the sigmoid function and can map the output in the range of 0–1, $h_{t-1}$ and $x_{t}$ are the previous hidden state and the input vector at time $t$, respectively. $W_{r}$ and $U_{r}$ are learnable weight matrices. In the same manner, the update gate $Z_{t}$ can be computed using
\begin{equation}\label{Z}
  {{Z}_{t}}={\rm \sigma} ({{W}_{xz}}{{x}_{t}}+{{W}_{hz}}{{h}_{t-1}}+{{b}_{z}}),
\end{equation}
where
\begin{equation}\label{ht'}
  {{\tilde{h}}_{t}}={\rm tanh}\left( {{W}_{xh}}{{x}_{t}}+{{W}_{hh}}\left( {{R}_{t}}\times {{h}_{t-1}} \right)+{{b}_{h}} \right),
\end{equation}
and
\begin{equation}\label{ht}
  {{h}_{t}}={{Z}_{t}}{{h}_{t-1}}+\left( 1-{{Z}_{t}} \right)\times {{\tilde{h}}_{t}}.
\end{equation}

According to Eq. (\ref{ht'}), the candidate hidden state ${\tilde{h}}_{t}$ forces to forget the previous hidden state $h_{t-1}$ and reset with the current input $x_{t}$ only when the reset gate $R_{t}$ is close to 0. Such mechanism effectively allows the hidden state to discard any information irrelevant in the future. In addition, the update gate $Z_{t}$ decides the amount of information carried over from the previous hidden state $h_{t-1}$ to the current hidden state $h_{t}$. These functions are similar to the memory unit in LSTM network and help RNNs to retain longterm information. Therefore, based on the GRUs model, we can effectively analyze long spectral sequences.

Since each hidden cell has a separate update gate and reset gate, they learn to capture dependencies in different wavelength frames. Those units learning to capture short-term dependencies tend to comprise frequently active reset gates, while those that capture longer-term dependencies comprise frequently active update gates. It implies that the model can not only capture long-term dependencies, but also remember these local important features. It is worth mentioning that CNNs are better at extracting spatial features, while RNNs are better at extracting sequence features. In some scenarios involving high-dimensional sequence information, such as electric field diagrams, these two advantages can be combined to achieve excellent performance.

\subsection{Constructing datasets}
\begin{figure}[h!]
\centering\includegraphics[width=13cm]{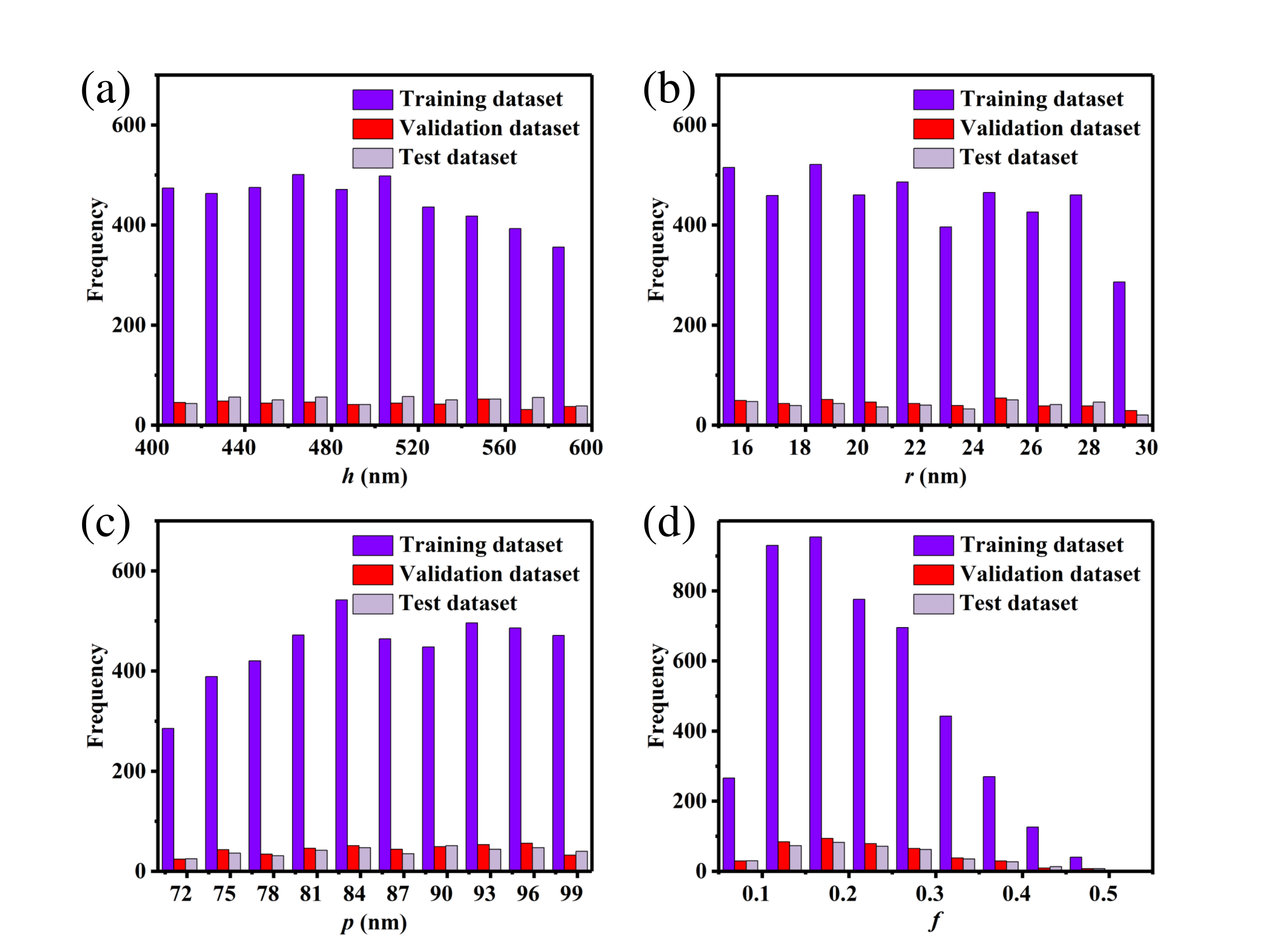}
\caption{Histograms of the three datasets. The training, validation, and test datasets contain 4500, 434, and 400 random parameter combination, respectively. (a) $h$. (b) $r$. (c) $p$. (d) $f$.}
\label{fig:3dataset}
\end{figure}

A dataset containing suitable samples is crucial for training any DL model. We prepare the dataset implemented by the finite difference time domain (FDTD) method. In the simulation, the moderate mesh grid is adopted for an effective tradeoff between accuracy, calculation time and memory requirements. The $p$-polarized plane wave is normal incident. Periodic boundary conditions are used in the x and y directions and the perfect matching layer conditions are adopted in the z direction. The refractive index of $\rm Al_{2}O_{3}$ and $\rm SiO_{2}$ are 1.75 and 1.44, respectively. The permittivity of gold can be described by the Drude model, $\varepsilon_{m}=1-\left[\omega_{p}^{2}/{\omega(\omega-i/\tau)}\right]$, where $\omega_{p}=2\pi \times 2.175\times 10^{15} $ $\rm s^{-1}$  is the plasma frequency of Au and $\tau$ is the relaxation time, $1/\tau=2\pi \times 6.5\times 10^{12} $  $\rm s^{-1}$ \cite{Au-liu2010infrared}. Owing to the slight effect of 5 nm Cr layer on the spectrum, the Cr layer is not considered in our simulation. The dataset is built by random uniform sampling of the geometric hyperspace, and the sampling intervals of the three parameters $r$, $p$, and $h$ are 0.2 nm, 0.2 nm and 1 nm, respectively. The ranges of parameter $r$, $p$, and $h$ are limited to 15-30 nm, 70-100 nm, and 400-600 nm, respectively. Every spectrum is resampled as 401 reflectance points between 800-2000 nm, and the length 401 is long enough to assess the model’s ability to process long-time sequence data. Based on the above settings, we collected 5334 sets of simulation results. The entire dataset is randomly divided into three subsets: 4500 data values for training, 434 data values for validation, and the remaining 400 data values for testing. Fig. \ref{fig:3dataset} shows the data distributions of the three datasets. There are two points worth noting. First, the data points are evenly distributed in each interval to ensure that the relationship between the structural parameters and the spectrum in each interval can be learned. Second, the proportional distribution of the three datasets in each interval is consistent, which is conducive for the robustness of the model.

\subsection{Inverse design spectrum}
In the inverse design module, the GRUs model is trained with the geometric structure ($f$/$r$/$h$) and spectrum of the NHMMs to accurately capture the spectrum-geometry relationships. According to our experience\cite{BPP2, HMM-Ultrabroadband}, $f$ has a greater impact on the spectrum than $p$, thus we choose $f$ instead of $p$ as the parameter. The well-trained GRUs model can resolve the dimensional parameters of NHMMs based on given or desired spectral information. In our study, the popular ML framework, TensorFlow 2.1, was used to build and train the GRUs. The architecture of the inverse design model is shown in Fig. \ref{fig:2GRU model}(b). A two-layer GRUs network is built to improve the model's understanding of spectral sequences. The units of the first hidden layer and the second hidden layer are 20 and 30, respectively. All timesteps are 401, which is consistent with the length of the spectrum sequence $R=({{r}_{{{\lambda }_{1}}}},\ldots \ldots ,{{r}_{{{\lambda }_{401}}}})$. All trainable parameters in this model are only 6255, which is close to two orders of magnitude less than the inverse design model based on FCNNs\cite{FCNN2}. Such a long spectrum requires $1.04\times10^{8}$ trainable parameters using the previous CNNs model\cite{CNN2}. Batch normalization is used between layers, which contribute to avoid overfitting and can speed up our model convergence. Considering that the height and radius values are relatively large in the output of the model, once outliers appear, it is not conducive to the training of the model. Therefore, Huber loss ($HL$) is selected as the loss function. $HL$ combines the best characteristics of mean absolute error ($mae$) and mean square error ($mse$), and reduces the impact of abnormal points, making training process more efficient. Besides, it is necessary to assign an appropriate weight to each parameter loss, which is beneficial for updating the gradient. Finally, the loss function can be expressed as: loss = 5$f_{HL}$ + $r_{HL}$ + 0.12$h_{HL}$, and is minimized by the Adam solver, which is computationally efficient and consumes less memory\cite{adam}. In addition, the learning rate is set as 0.055 initially with a decay of 4\% every 5 epochs and the trainset is input into the model with batch sizes of 1200.
\begin{figure}[h!]
\centering\includegraphics[width=13cm]{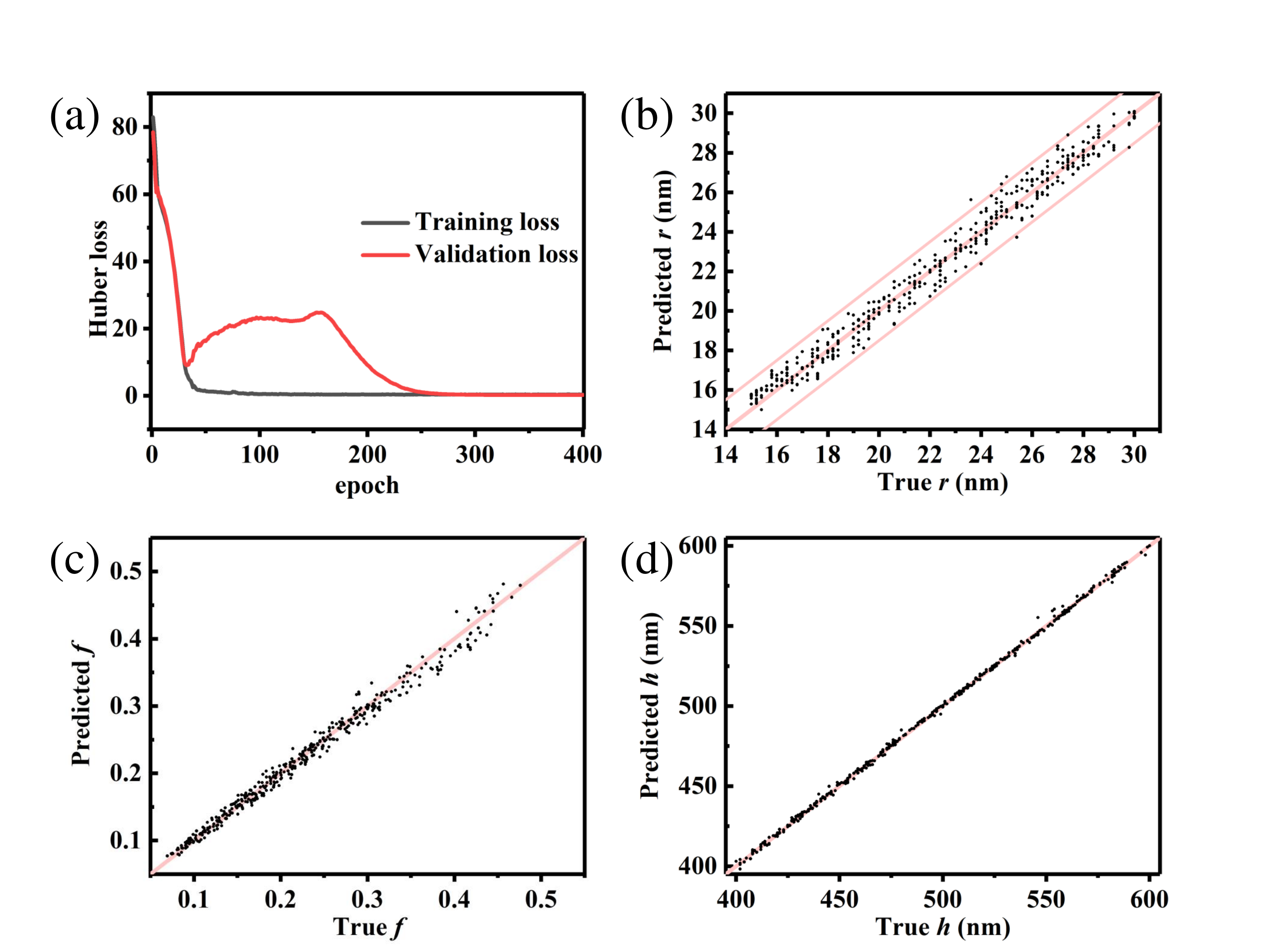}
\caption{Performance of GRUs inverse design model. (a) Huber loss, black (red) represents training (validation) loss. After 400 epochs, the training loss and validation loss decay to 0.3205 and 0.2802, respectively. (b), (c) and (d) Scatter plot of true (red line) and predicted (black dots) values of $r$, $f$, and $h$.}
\label{fig:4loss}
\end{figure}

The training performance of the GRUs inverse design model is presented in Fig. \ref{fig:4loss}. After 400 epochs, the values of training and validation losses are 0.3205 and 0.2802, respectively. The red curve represents the validation loss, which first decreases, then increases, and finally decays to a stable value, indicating that there is a process of resisting overfitting. Then, the 400 spectrums in the test set are fed to the trained model and the predicted parameters $f$/$r$/$h$ are compared with the true parameters generating the spectrum. It is worth mentioning that the model only took a few seconds to obtain the structural parameters corresponding to 400 sets of spectrum. Figs. \ref{fig:4loss}(b)-(d) show the comparisons. Thus, the $mae$ of the three parameters ($f$/$r$/$h$) are only 0.0083, 0.4802 nm, and 0.9966 nm, respectively. The mean relative error ($mre$) of the three parameters ($f$/$r$/$h$) are 3.79\%, 2.23\%, and 0.2\%. These important results show that the structural parameters obtained by the GRUs inverse design model are appreciably accurate. As shown in Fig. \ref{fig:4loss}(b), the three red lines are plotted as guidelines, denoting $y = x$ and $y =x \pm 1.5$. The predicted $r$ follow the trend of $y = x$ closely and 98\% of the predicted values fall within the boundaries of $y=x \pm 1.5$. As shown in Fig. \ref{fig:4loss}(c), the predicted $f$ are close to the real $f$ in the range of smaller $f$. However, the prediction values are slightly farther for larger $f$. This problem can be attributed to the relatively less number of the large $f$ values in the dataset, thus obtaining no perfect match between the input and $f$ in the model training process, which leads to the deterioration of the prediction performance for large $f$ values. Fig. \ref{fig:4loss}(d) shows that the predicted $h$ are tightly distributed on and near the red line, showing good consistency between the true and predicted values.

\begin{figure}[h!]
\centering\includegraphics[width=13cm]{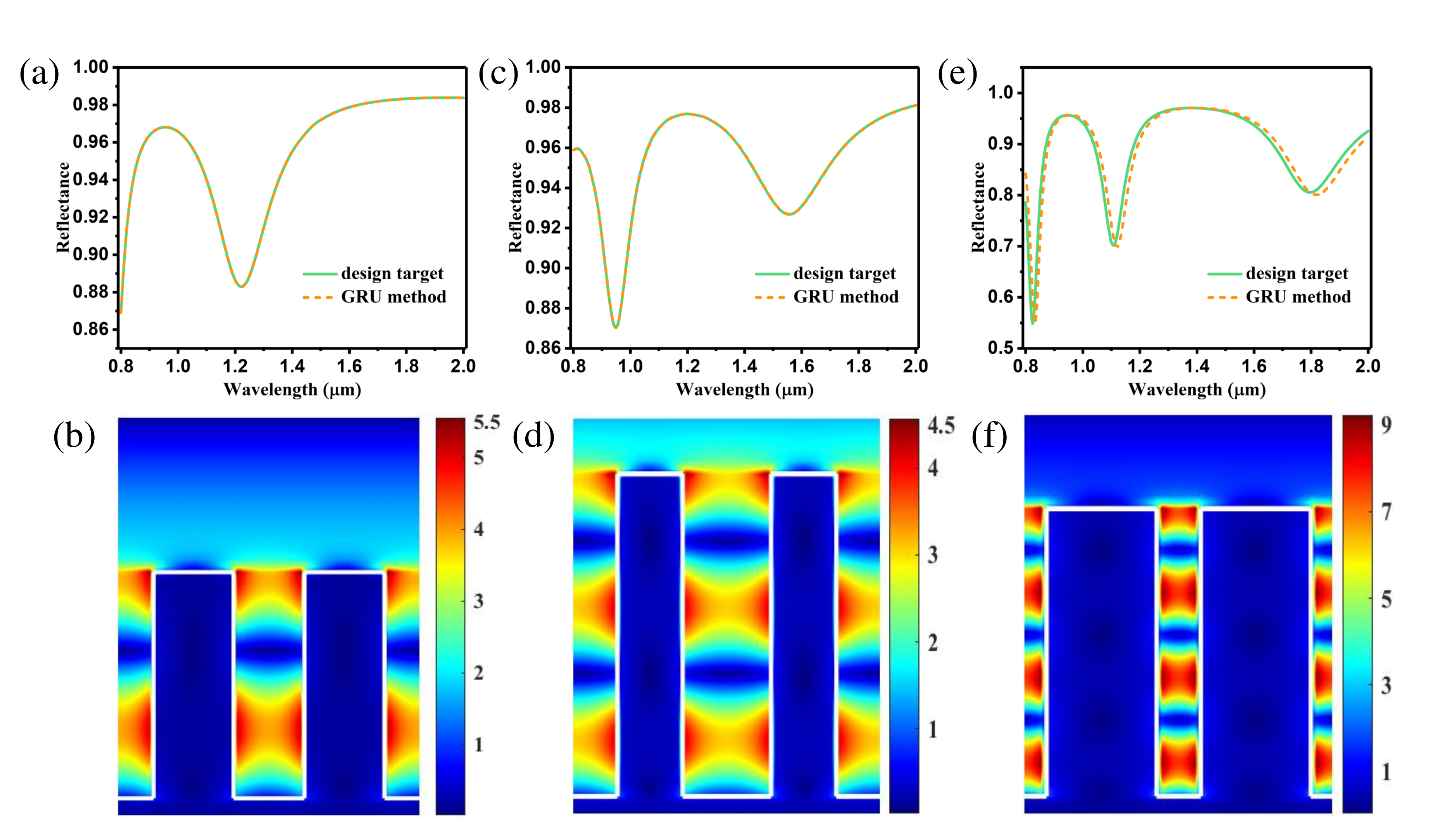}
\caption{Comparison between the target spectrum and the one simulated from the GRU method predicted parameters. (a), (c) and (e) show 1, 2, and 3 resonances in the 800-2000 nm band respectively. (b), (d) and (f) represent electric field intensity distributions corresponding to the resonance at 1223 nm in (a), 947 nm in (c), and 884 nm in (e), respectively.}
\label{fig:5E}
\end{figure}

In order to further assess the performance of our model for spectrum design, three target response with one, two, and three resonance peaks are fed to the GRUs model. Subsequently, FDTD simulations are implemented using the predicted structural parameters to obtain corresponding reflection spectrum. Fig. \ref{fig:5E}(a), (c), and (d) depict desired (green solid) and predicted (orange dashed) spectrum of three NHMMs with dimensions $f$/$r$/$h$ of 0.22/22.4 nm/402 nm, 0.14/19.6 nm/576 nm, and 0.40/26.8 nm/501 nm, respectively. Mode characteristics of NHMMs can be reflected in their spectrums from both the GRUs prediction and FDTD simulation. A visible redshift of the resonance mode can be found owing to the increase of filling fraction or height. Furthermore, significant fitting between the desired and forecasted spectrum reveals the strong confidence of GRUs designed optical parameters.

To investigate the origin of such plasmonic resonance of NHMMs, we plot the electric field intensity distributions on resonance in Figs. \ref{fig:5E}(b/d/f).
The spatial confinement of plasmon-polaritons associated with cavity resonances is the cause of the periodic mode structure in the NHHMs. Figs. \ref{fig:5E}(b/d/f) show the electric fields confined in adjacent nanorods, indicating that different orders of bulk plasmon-polaritons modes are excited\cite{BPP1,BPP2}. The difference between Figs. \ref{fig:5E}(d) and Fig. \ref{fig:5E}(f) indicates that a large filling fraction $f$ increases the effective refractive index of the cavity, resulting in more modes appearing in the reflection spectrum. Similarly, comparing Fig. \ref{fig:5E}(b) and (d), the increase in height or cavity length causes more modes to appear in the reflection spectrum.

\subsection{Predict unknown spectrum}
\begin{figure}[h!]
\centering\includegraphics[width=12cm]{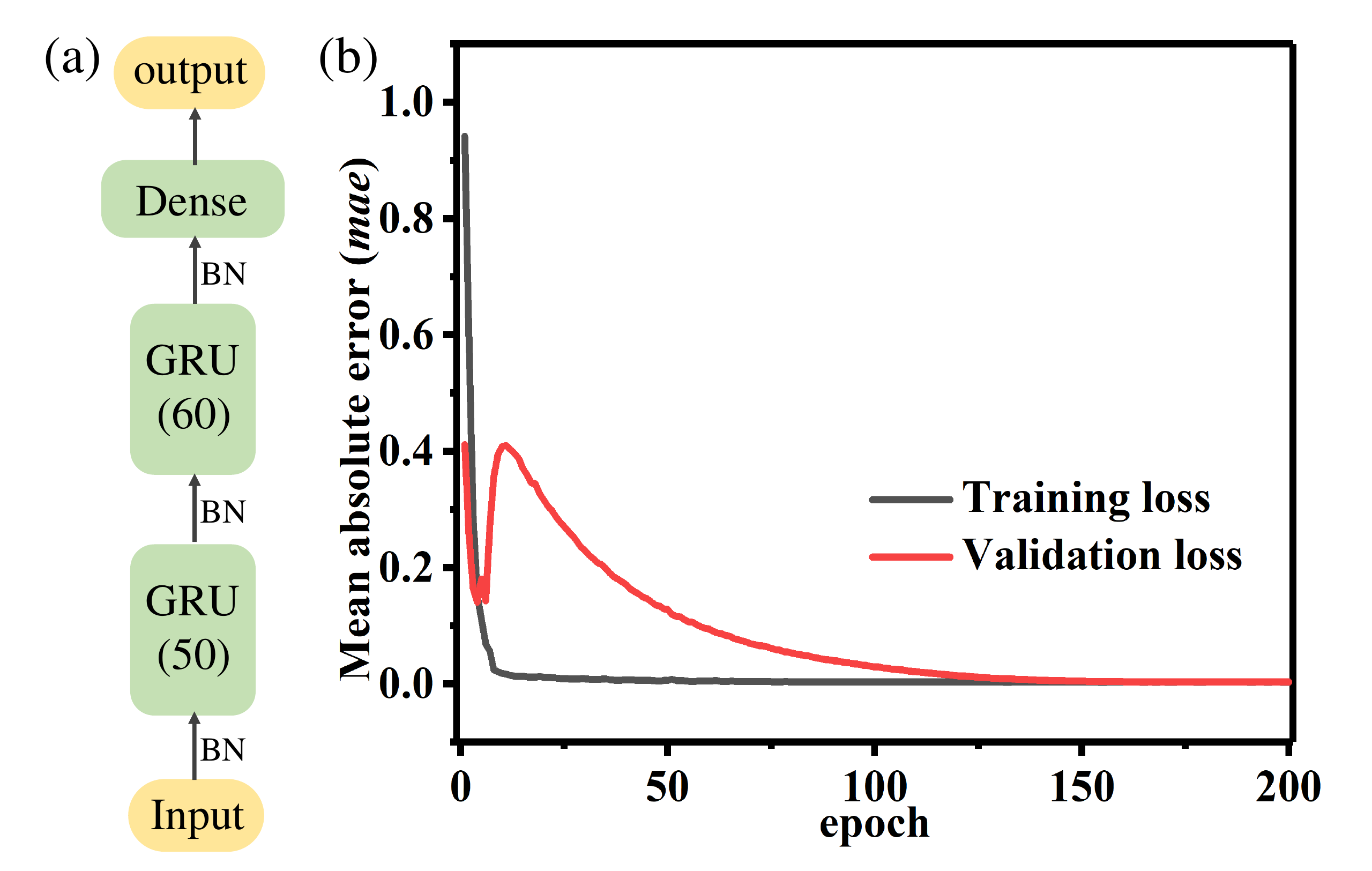}
\caption{(a) Schematic of the GRUs prediction model. The input of the model is 301 reflectance points in the range of 800-1700 nm, and the output is 100 reflectance points in the range of 1703-2000 nm. (b) The loss function $mae$ of training and validation. After 400 epochs, the training loss and validation loss decay to 0.0.003 and 0.0031, respectively.}
\label{fig:6GRU_predict_model}
\end{figure}
In the previous section, we have demonstrated the accuracy and efficiency of inverse design using GRUs. In addition, the characteristics of the GRUs network can be further used for spectral prediction. Given the success of GRUs method to predict unknown sequence data, such as stock trends and weather prediction, it makes sense as a plausible candidate solution to predicting the spectra of nanophotonic devices. Next, we demonstrate the use of the GRUs network to build a prediction model to predict the spectrum of the 1703-2000 nm band based on the spectrum of the 800-1700 nm band. Still based on the previously prepared dataset, we only exploit the spectral information. In the predictive model, 301 reflectance data points in the range of 800 nm to 1700 nm are used as input, and the remaining 100 reflectance data points are used as label. The frame of GRUs prediction model is shown in Fig. \ref{fig:6GRU_predict_model}(a). Compared to the inverse design model, the architecture is generally similar, however, the units of the prediction model in the first and second hidden layers are 50 and 60, respectively, and the number of timesteps are 301. $Mae$ is chosen as the loss function. The learning rate is set as 0.015 initially with a decay of 4\% every 5 epochs. The trainset is input into this model with batch sizes of 800.
Without large effort, we got an accurate result. As shown in Fig. \ref{fig:6GRU_predict_model}(b), after 200 epochs, the values of training and validation losses decay to 0.0030 and 0.0031, respectively. Afterwards, the test set is fed to the trained prediction model to assess the generalization ability. The evaluation indices $mae$, $mse$, and $mre$ are $2.8\times10^{-3}$, $8.8\times10^{-5}$ and 0.32\%, respectively. These results strongly demonstrate the effectiveness of the prediction model.

\begin{figure}[h!]
\centering\includegraphics[width=14cm]{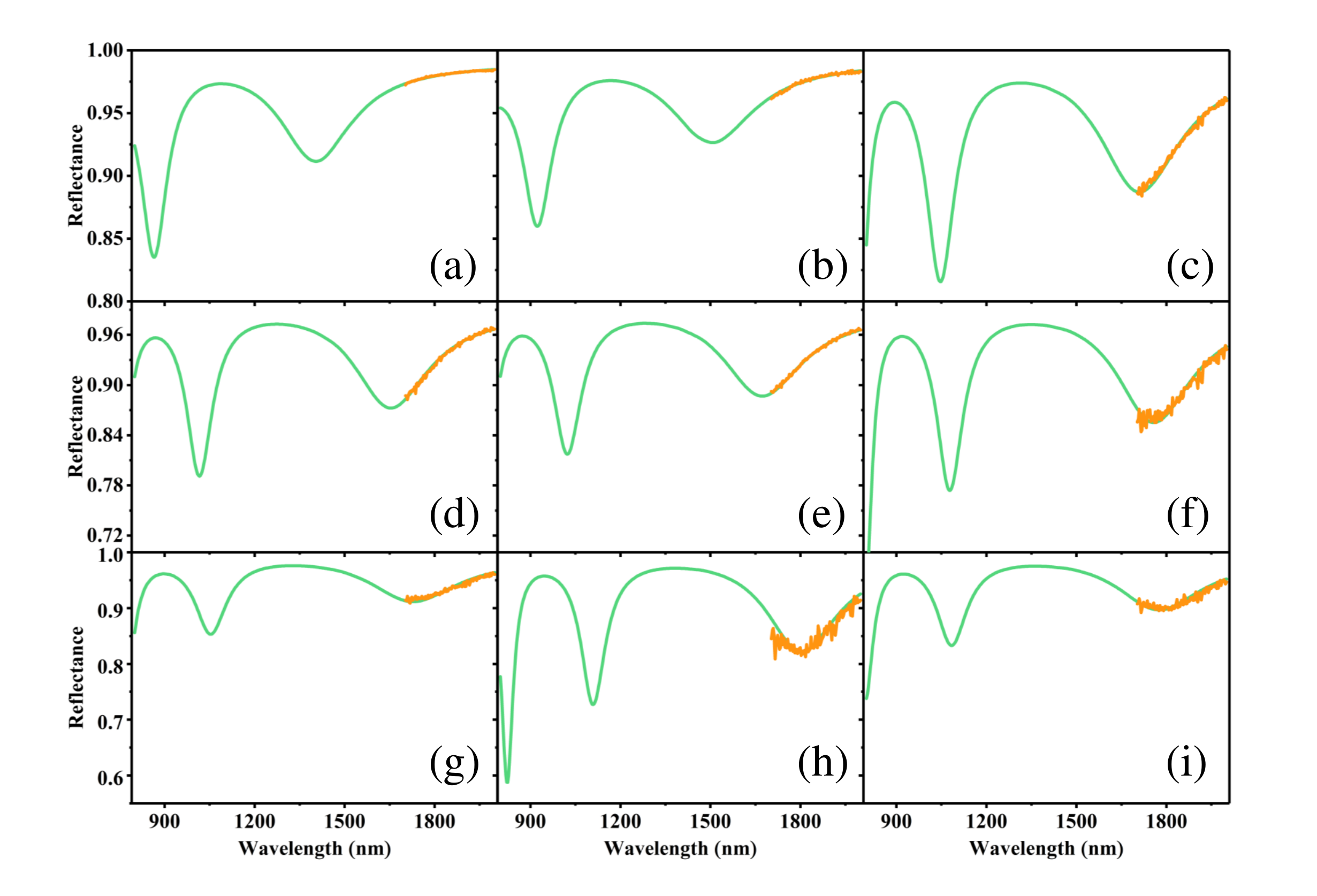}
\caption{Performance of the GRUs prediction model. (a)-(i) are 9 different spectrums, including flat, rising, and concave curves. The green curve in the 800 nm-2000 nm range represents the true spectrum. The orange curve in the range of 1703 nm-2000 nm represents the predicted spectrum.}
\label{fig:7GRU_predict}
\end{figure}

To demonstrate the ability of our model to predict spectrums of various curves, nine different spectrums from the test set are selected as display. As shown in Fig. \ref{fig:7GRU_predict}, the predicted orange spectrum and the green spectrum overlap well, showing good consistency between the predicted and true reflection values. In particular, for these flat spectra shown in Fig. \ref{fig:7GRU_predict}(a) and (b), the predicted values are almost identical to the true values. For these spectra showing an upward trend, as shown in Fig. \ref{fig:7GRU_predict}(c)-(e), the predicted values and the true values are also in good agreement. For more complex spectra, such as the inflection points shown in Fig. \ref{fig:7GRU_predict}(f)-(i), despite some glitches, the predicted values are very close to the true values. If we further enrich those spectral data with complex curves like Fig. \ref{fig:7GRU_predict}(f)-(i) in the 1703-2000 nm band in the trainset, the performance of the prediction model can be definitely improved.

We demonstrated the effectiveness of inverse design and spectrum prediction of nanophotonic devices using RNNs. Our design method possesses two distinct advantages. First, when understanding a spectrum, it is not enough to understand each reflectance point in isolation. A better way is to process the entire spectrum sequence connected by these reflectance points. Therefore, RNNs with memory or feedback loops allow solving inverse design tasks related to time series data efficiently. Second, the trainable parameters of the RNNs based model are fewer than those of the FCNNs or CNNs based model. Moreover, the RNNs model has only few hyperparameters and one can adjust to the appropriate parameters without great effort, saving a lot of time.
\section{Conclusion}
In summary, we demonstrated a novel DL method using GRUs as an outstanding method to achieve inverse design and spectrum prediction for nanophotonic devices. GRUs have memory and feedback, thus, it performs better on recognizing spectral data with sequence characteristics. We used the widely studied NHMMs as an example to demonstrate the effectiveness of this method. First, we constructed an inverse design model. The reflection spectrums designed by GRUs are approximate to the FDTD-simulated results with high precision. It is important to note that the model has learned the complex physical relationship about how the structural parameters affect the spectrum. Second, we established a spectral prediction model, which is capable of predicting unknown spectrum based on known spectrum with only 0.32\% mean relative error. More importantly, it can accurately predict various curves. Furthermore, the model can convert complex physics problems into calculation problems with fewer parameters compared to FCNNs and CNNs, and less time is required to tune hyperparameters. In the future, this method can be combined with CNNs to process high-dimensional sequence data like electric field distribution, holograms, etc. Therefore, the GRUs model with high precision, low computational cost, and good capability of processing spectrum was demonstrated to show illimitable potential for nanophotonic design.
\section*{Funding}
National Natural Science Foundation of China (NSFC) (61775064); Fundamental Research Funds for the Central Universities (HUST: 2016YXMS024).

\section*{Acknowledgments}
The author Ruoqin Yan (RQYAN) expresses his deepest gratitude to his PhD advisor Tao Wang for providing guidance during this project.

\section*{Disclosures}
The authors declare no conflicts of interest.
\bibliography{references}

\end{document}